\documentclass[12pt]{article}
\usepackage[a4paper, margin=1.5in]{geometry}
\usepackage{amsmath}
\usepackage{amssymb}
\usepackage{array}
\usepackage{natbib}
\usepackage{graphicx}
\usepackage{doi}
\usepackage{layout}

\numberwithin{equation}{section}
\DeclareMathOperator*{\argmax}{arg\,max}

\newtheorem{myth}{Theorem}

\newenvironment{proof}{\vspace{-0.1cm}\noindent\textbf{Proof:}}{$\Box$\\}

\begin{document}

	\title{Proportional resource allocation in dynamic $n$-player Blotto games\thanks{First version was entitled ``Multi-Battle $n$-Player Dynamic Contests'' (2018). We are grateful to Jean-Jacques Herings, Arkadi Predtetchinski, and J\'{a}nos Flesch without whose feedbacks it would not be possible to complete this paper. We would like to thank Steven Brams, Kang Rong, Jaideep Roy, Christian Seel, C\'{e}dric Wasser, and audiences at Maastricht University, HEC Montreal, and Games and Contests Workshop at Wageningen, King's College London, Durham University, Sabanci University, the 7th Annual Conference on ``Contests: Theory and Evidence,'' and 2022 Conference on Mechanism and Institution Design for their valuable comments. }}
	\author{ Nejat Anbarci\thanks{%
			Department of Economics and Finance, Durham University, Durham DH1 3LB, UK. Nejat.anbarci@durham.ac.uk} \and Kutay Cingiz\thanks{%
			Agricultural Economics and Rural Policy Group, Wageningen University, Hollandsweg 1, 6706KN Wageningen, The Netherlands. Kutay.cingiz@wur.nl} \and Mehmet S.
		Ismail\thanks{%
			Department of Political Economy, King's College London, London, WC2R 2LS, UK. ORCID: 0000-0002-3232-0776. mehmet.s.ismail@gmail.com}}
	
	\date{This version: July 2022}
	
	\maketitle
	\begin{abstract}
		
		A variety of social, economic, and political interactions have long been modelled after Blotto games. In this paper, we introduce a general model of dynamic $n$-player Blotto contests. The players have asymmetric resources, and the battlefield prizes are not necessarily homogeneous. Each player's probability of winning the prize in a battlefield is governed by a contest success function and players' resource allocation on that battlefield. We show that there exists a subgame perfect equilibrium in which players allocate their resources proportional to the battlefield prizes for every history. This result is robust to exogenous resource shocks throughout the game.   
		
		
	\end{abstract}
	
	\thispagestyle{empty}
	\pagenumbering{arabic}
	\setcounter{page}{0}
	\newpage
	

	\section{Introduction} 
	
	There are many social, economic, and political interactions that can be modelled as contests. Examples include rent-seeking, political campaigns, sports competitions, litigation, lobbying, and wars, among others. Traditionally, the first contest model of ``Colonel Blotto'' game was introduced by Borel (1921). A Colonel Blotto game is a two-person static game in which each player allocates a limited resource over a number of identical ``battlefields.'' Fast forward, the literature on contests is now enormous, and a Blotto contest denotes any contest in which two or more players allocate a limited resource over some battlefields.
	
	In this paper, we introduce a dynamic multi-battle $n$-player Blotto games as follows. The players each have possibly asymmetric resources and the battlefields each have possibly a distinct prize. Each player's probability of winning the prize in a battlefield is governed by a contest success function (CSF) and players' resource allocation on that battlefield. In our benchmark model, each player starts the dynamic contest with a limited budget and distributes this budget over a finite number of battlefields. Since the battles take place in a sequential order, players can condition their strategies on the outcomes of previous battles. At time $t$, players choose simultaneously their allocation on battlefield $t$ to win $x^{t}>0$, which is the prize of the battlefield. The winner of the prize in a battlefield is determined by a contest success function satisfying the axioms (A1-A6) of Skaperdas (1996). The winner and the resulting resource allocations are revealed to every player before the next battle takes place. The players maximize the total expected prizes in this dynamic game. We also introduce an extension of our model to the case in which the initial resources of the players may be subject to an exogenous shock and hence can change throughout the game.
	
	Studying a static, simultaneous-move model of
	resource allocation in U.S. presidential campaigns in a prominent paper,
	Brams and Davis (1974) highlighted the concepts of `population of states' as
	well as `population proportionality' in campaign resource allocation.%
	\footnote{%
		As noted by Brams and Davis (1974), the population of a state need not
		exactly reflect the proportion of the voting-age population who are
		registered and actually vote in a presidential election.} Brams and Davis
	(1974, p. 113) showed that populous states receive disproportionately more
	investments with regards to their population. More specifically, the
	winner-take-all feature of the Electoral College---i.e., that the
	popular-vote winner in each battle wins all the electoral votes of that
	battle---induces candidates to allocate campaign resources roughly in
	proportion to the $3/2$'s power of the electoral votes of each state. The
	question of why some small states ``punch above their
	weight''---i.e., attract attention and resources more than
	proportional to their weight---in political campaigns has been puzzling
	researchers (for an analysis in a non-Blotto setting, paying attention to
	``momentum'', see, e.g., Klumpp and Polborn, 2006).
	
	The proportional allocation of resources is generally considered to be a benchmark in resource distribution games, and especially in Blotto contests it is not only one of the prominent strategies but arguably the most salient heuristic. In a symmetric experimental Blotto game, Arad and Rubinstein (2012) consider the equal distribution of resources as level-0 behavior; it also seems to be the first strategy that comes to mind because of the low response time associated with it. (For level-$k$ reasoning, see Stahl, 1993, and Nagel, 1995.) As we discuss next, we find that this prominent heuristic is an equilibrium outcome in our setting.\footnote{Both in the original Blotto game and in the one considered by Arad and Rubinstein (2012), all battlefields have the same prize. With heterogeneous battlefields that we consider, equal distribution of resources corresponds to a distribution of resources proportional to different battlefield prizes.}
	
	Our solution concept is subgame-perfect equilibrium. We find that the strategy profile in which players allocate their resources proportional to the prizes of the battlefields at
	every history is a subgame perfect equilibrium (see Theorem~\ref{thm2} and Theorem~\ref{thm3}). This overall result does not depend on the number of players, asymmetry in the resources, exogenous shocks to resources, number or the prizes of battlefields, or the type of contest success functions satisfying the aforementioned axioms.
	
	\section{Relevant Literature}
	\label{sec:literature}
	
	Our paper contributes to the literature on dynamic contests and campaign resource allocation in sequential elections. This brief section summarizes related work apart from the ones mentioned in the Introduction. 
	
	Friedman (1958) first shows that proportional allocation is a Nash equilibrium in static $2$-player Blotto contests with Tullock CSF (see equation~\ref{eq:Tullock} in section~\ref{sec:expected_value}). Osorio (2013) extends Friedman's result to the case in which battlefield prizes are asymmetric. Duffy and Matros (2015) extend Friedman's result to static $n$-player Blotto contests with Tullock CSF. 
		
	In a recent closely related paper, Klumpp, Konrad, and Solomon (2019) consider dynamic Blotto games where two players fight in odd number of battlefields, which are identical.\footnote{Colonel Blotto games were first introduced by Borel (1921). Among others, recent contributions to Blotto games include Roberson (2006), Kvasov (2007), and Rinott, Scarsini, and Yu (2012). There is also a huge literature on non-Blotto contests initiated by Tullock (1967; 1974), and see also, e.g., Krueger (1974). The early literature on non-Blotto contests are motivated by rent-seeking.} The player who wins the majority of battles wins the game. Accordingly, they show that under general contest success functions players allocate their resources evenly (i.e., proportionally by default) across battlefields in all subgame perfect equilibria, one of which is in pure strategies.\footnote{For a discussion of dynamics in contests, see Konrad (2009).} A more recent follow-up paper by Li and Zheng (2021) study Klumpp et al.'s even-split strategy in a more general setting.
	
	Acharya, Grillo, Sugaya, and Turkel's (2021) paper builds on Klumpp et al.'s (2019) by studying dynamic electoral campaigns as dynamic contests with two players whose `relative popularity' changes over time. Acharya et al.'s contests are also of Blotto type in the sense that the two players have fixed resources to allocate. However, their model differs from Blotto contests in that players' investments affect the evolution of popularity via a Brownian motion. In their setting, Acharya et al. also confirm the even-split result of Klumpp et al. (2019).

	Sela and Erez (2013) studied a two-player dynamic Tullock contest, in which each player maximizes the sum of the expected payoffs (similar to our setting) for all districts. Suppose that the prizes of the battlefields are equal across the stages and that for each resource unit that a player allocates, they lose $0\leq \alpha \leq 1$ units of resource from their budget.
	Then Sela and Erez (2013) identified a subgame prefect equilibrium such that the players'
	resource allocations are weakly decreasing over the stages. Duffy and Matros
	(2015) studied static contests (stochastic asymmetric Blotto games) in up to four battlefields with two players having asymmetric yet similar budgets and generalizing Lake's (1979) paper, which we discuss below.\footnote{For experimental results on Blotto games see, e.g., Deck and Sheremeta (2012), Montero, Possajennikov, Sefton, and Turocy (2016), Duffy and Matros (2017) and the references therein.} In a similar setting, Deck, Sarangi, and Wiser (2017) have recently studied symmetric static contests with two players who do not have budget constraints. They identified the Nash equilibrium of the symmetric game (Electoral College).
	
	In another static presidential campaign model, Lake (1979) argued that one would need to assume that the candidates maximize only their probability of winning the election, i.e., one would simply try to receive a majority of electoral votes, instead of complying with Brams and Davis' (1973, 1974) assumption that they maximize their expected electoral vote. Nevertheless, Lake's (1979) main result echoes Brams and Davis' (1974) impossibility of population proportionality result in that in Lake's model too it turns out that presidential candidates find it optimal to spend a disproportionately large amount of their funds in the larger states.\footnote{Resource allocation frameworks are often used in modeling electoral competition; see, e.g., Laslier and Picard (2002), Duggan (2007), Barelli, Govindan, and Wilson (2014), Thomas (2017), and the references therein.}
	
	Harris and Vickers (1985) construed a patent race as a multi-battle contest, in which two players alternate in expending resources in a sequence of single battles. These battles or sub-contests serve as the components of the overall R\&D contest. Just like in a singles tennis match, the player who is first to win a given number of battles wins the contest, by obtaining the patent.\footnote{In the PGA Tour, which brings professional male golfers together to play in a number of tournaments each year (LPGA does so for female golfers), each tournament consists of multiple battles in that golfers attempt to minimize the total number of shots they take across 72 holes.}
	
	Additional work on dynamic resource allocation contests is as follows.
	Dziubi\'nski, Goyal and Minarsch (2017) have recently studied multi-battle
	dynamic contests on networks in which neighboring `kingdoms' battle in a
	sequential order. Hinnosaar (2018) chracterizes equilibria of sequential contests in which efforts are exerted sequentially to win a (single-battle) contest. Ewerhart and Teichgr{\"a}ber (2019) study multi-battle dynamic non-Blotto contests and show the existence of a unique symmetric Markov perfect equilibrium. In a two-player and two-stage campaign resource allocation
	game, Kovenock and Roberson (2009) characterized the unique subgame perfect
	equilibrium. In a two-player best-of-three multi-battle dynamic contest, Konrad (2018) anaylzes resource carryover effects between the battles. Brams and Davis (1982) examined a model of resource allocation in the U.S. presidential primaries to study the effects of momentum transfer from one primary to another. As alluded before, Klumpp and Polborn (2006) also focused on momentum issues; they considered a two-player model in which an early primary victory increases the likelihood of victory for one player and creates an asymmetry in campaign spending, which in turn magnifies the player's advantage. This asymmetry of campaign spending generates a momentum which can propel an early winner to the overall victory. Strumpf (2002), on the other hand, discussed a countervailing force to momentum, which favors later winners.
	
	\section{Model} 
	\label{sec:model}
	
	We consider dynamic Blotto contests where there are $m$ heterogeneous battlefields with a predetermined \emph{sequential} order, indexed by $t= 1, 2, ..., m$, and $n$ players, indexed by $i = 1, 2, ..., n$. Players have possibly asymmetric (sunk) budgets: Each player $i$ has a budget $W_{i}\geq 0$ that he or she can allocate over the battlefields. The \textit{prize} of each battlefield $t$ is denoted by $x^{t}>0$. Each time period $t$, the battle at $t$ takes place, and each player $i$ \emph{simultaneously} chooses a pure action (allocation) denoted by $w_{i}^{t}$ which is smaller than or equal to the budget, $W_{i}$, minus the already spent allocation by player $i$ until battle $t$. Given the chosen actions in battle $t$, $w^{t}:=(w_{1}^{t},\ldots ,w_{n}^{t})$, the probability of player $i$ winning battle $t$ is defined by a contest success function, which has the following form.
	
	\begin{equation}
		p_i ^t(w^t)=\begin{cases}
			\frac{f(w_{i}^{t})}{ \sum_{j}f(w_{j}^{t})} & \textrm{if}~\sum_{j}w_{j}^{t}>0 \\
			\frac{1}{n} & \textrm{if}~\sum_{j}w_{j}^{t}=0,
		\end{cases}
	\end{equation}
	where $f(.)$ satisfies Skaperdas's (1996) axioms (A1-A6), which characterizes a wide range of contest success functions used in the literature. More specifically, it is of the following form: $f(w_{j}^{t})=\beta (w_{j}^{t})^\alpha$ for some $\alpha>0$ and $\beta>0$. 
	
	To avoid trivial cases, we assume that for any $t$,  $x^t < \sum_{t' \neq t}x^{t'}$, that is, there is no ``dictatorial'' battlefield. Let $x_i ^{t}$ be the prize player $i$ wins at battle $t$, which is $x^t$ with probability $p_i ^t(w^t)$ or 0 with probability $1-p_i ^t(w^t)$.\par
	The set of histories of length $t$ is denoted by $H^t$. A history of length $t \geq 1$ is a sequence
	\begin{equation}
		h^t :=(((w_1 ^{1},x_1 ^1),\ldots,(w_n ^{1},x_n ^1)),\dots,((w_1 ^{t},x_1 ^t),\ldots,(w_n ^{t},x_n ^t)))
	\end{equation}
	satisfying the following conditions
	\begin{itemize}
		\item[(i)] For each $1 \leq i \leq n$ and for each $1 \leq t'\leq t$,  $w_i^{t'} \in [0,W_i-\sum_{j< t'}w_i ^{j}]$. \par
		\item[(ii)] For each battle $t'\leq t$, there exists a unique player $i$ such that $x_i^{t'} = x^{t'}$ and for all $j \neq i$, $x_j^{t'}=0$.
	\end{itemize}
	The first property states that each action at any given battle $t$ is bounded by the budget set which diminishes after each action taken in previous battles. The second property states that the battles each have winner-take-all structure.
	
	The history $H ^0$ consists of only the empty sequence $\o$. Let $H=H^0 \cup H^1 \cup \ldots \cup H^m$. Note that, the history $h^{t-1}$ is presented to all players at time $t$. There is a subset $\overline{H^t} \subset H$ consisting of histories of length $t$ where the game comes to an end at battle $t$. We call $\overline{H^t}$ the set of terminal histories of length $t$. If the game has not ended before battle $m$ then the game ends at battle $m$. We will specify terminal histories in detail later on.\par
	The remaining budget of player $i$ after history $h^t\in H$ is defined as $B_i(h^t)= W_i -\sum_{j\leq t} w_i ^{j}$ where for every $j\leq t$, $w_i ^{j}$ is a realized spending of history $h^{t}$. The realized winning schedule of a given history $h^t \in H$, denoted by $V(h^t)$, is the sequence of players that won the battles at battlefields $1,\ldots,t$. Thus $V(h^t)\in \{1,\ldots,n\}^{t}$. For example, if $h^3=(((w_1 ^{1},x_1 ^1),(w_2 ^{1},0)),((w_1 ^{2},0),(w_2 ^{2},x_2 ^2)),((w_1 ^{3},0),(w_2 ^{3},x_2 ^3)))$ in a two-player dynamic contest with $m>3$ battlefields, then $V(h^3)=(1,2,2)$.\par
	For player $i$, a pure strategy $\sigma_i$ is a sequence of $\sigma_i^{t}$'s such that for each $t$, $\sigma_i^t$ assigns, to every $h^{t-1} \in H^{t-1}$, allocation $\sigma_i^t(h^{t-1}) \in [0,B_i(h^{t-1})]$. A pure strategy profile is denoted by $\sigma=(\sigma_i)_{i\leq n}$. The set of pure strategies of player $i \leq n$ is denoted by $\Sigma_i $ and the set of pure strategy profiles by $\Sigma =  \times_{i \leq n} \Sigma_i $. For any $\sigma \in \Sigma$, let $(\sigma| h)=((\sigma_1| h),\ldots, (\sigma_n| h))$ denote the strategy profile induced by $\sigma$ in the subgame starting from history $h$.\par
	
	Players maximize the expected payoff which is defined as the sum of expected battlefield prizes. So, the terminal histories are exactly the histories with length $m$. The set of terminal histories is denoted by $\overline{H^m}$, which is equal to $H^m$. 
	
	For any $\bar{h}^m \in \overline{H^m}$, player $i$ receives a payoff equal to
	\begin{equation}
		\bar{u}_{i}(\bar{h}^m)=\sum_{t\leq m} x_i ^t,
	\end{equation} where $\bar{h}^m:=(((w_1 ^{1},x_1 ^1),\ldots,(w_n ^{1},x_n ^1)),\dots,((w_1 ^{m},x_1 ^m),\ldots,(w_n ^{m},x_n ^m)))$.\par
	The set of terminal histories induced by a strategy profile $\sigma$ conditional on reaching history $h$ is denoted by $\rho(\sigma |h)$, which is a subset of $\overline{H^m}$. The payoff for player $i \leq n$ induced by a pure strategy profile $\sigma \in \Sigma$ at any $h^t \in H^t$ is
	\begin{align}
		v_i(\sigma|h^t)= \sum_{\bar{h}^m\in \rho(\sigma|h^t)} q(\sigma,\bar{h}^m|h^t)\bar{u}_i(\bar{h}^m).
	\end{align}
	
	Our solution concept is subgame perfect equilibrium in pure strategies.
	
	\noindent\textbf{Subgame perfect equilibrium:} A pure strategy profile $\sigma \in \Sigma$ is a \textit{subgame~perfect} \textit{equilibrium} if for every battle $t\leq m$, for every history $ h \in H^t$, for every player $i \leq n,$ and for every strategy $\sigma_{i}' \in \Sigma_i$
	\begin{equation*}
		v_{i}(\sigma|h) \geq v_{i}(\sigma_{-i},\sigma_{i}'|h). \label{spne}
	\end{equation*}
	A strategy profile $\sigma \in \Sigma $ is a subgame perfect equilibrium if and only if for every $ h \in H, $ $ \sigma $ induces an equilibrium in the subgame starting with history $h$. 
	
	\section{Dynamic Blotto games with Tullock contest success function}
	\label{sec:expected_value}
	
	In this section, we first define the well-known Tullock contest success function. 
	
	\begin{equation}
	\label{eq:Tullock}
		p_i ^t(w^t)=\begin{cases}
			\frac{w_{i}^{t}}{ \sum_{j}w_{j}^{t}} & \textrm{if}~\sum_{j}w_{j}^{t}>0 \\
			\frac{1}{n} & \textrm{if}~\sum_{j}w_{j}^{t}=0.
		\end{cases}
	\end{equation}
	
	A dynamic Tullock contest is a dynamic contest in which the contest success functions are of the Tullock variety. 
	
	The following theorem provides our first main result in which we show that a subgame perfect equilibrium in ``proportional strategies'' exists in every dynamic Tullock contest. Note that in the following section we generalize our model and extend this result to more general contest success functions.
	
	\begin{myth} [Existence and characterization: Tullock contests]
		\label{thm2} For any $n$-player dynamic Tullock contest, there exists a subgame perfect equilibrium, $\sigma$, which is given as follows. For any $t$, for any nonterminal history $h^{t-1}\in H-\overline{H}$, and for any player $i$, let
		\begin{equation}
		\sigma_i^{t}(h^{t-1})= B_i(h^{t-1}) \frac{x^{t}}{x^{t}+\ldots+x^{m}}.
		\end{equation}
		
	\end{myth}
	\begin{proof}
		We show that the proportional strategy profile $\sigma=(\sigma_{i})_{i \leq n}$ given above is robust to one-shot deviations, which implies that $\sigma$ is a subgame perfect equilibirum. That is, any player $i$ at any nonterminal history $h^t$ can not improve his payoff by changing $\sigma_i^t$, given that all other players, $j\neq i$, follow the proportional strategy. If player $i$ switches to a strategy $\bar{\sigma}_i=(\bar{\sigma}_i^{t+1},\sigma_i^{t+2},\ldots,\sigma_i^m)$ after history $h^t$ such that $\bar{\sigma}_i^{t+1}(h^t)\neq \sigma_i^{t+1}(h^t)$, then the expected prize player $i$  wins after battle $t$ given the history $h^t$ is denoted as $v_{i,t+1}(\bar{\sigma}_i,\sigma_{-i}|h^t)$, which satisfies 
		\begin{equation}
			v_{i,t+1}(\bar{\sigma}_i,\sigma_{-i}|h^t)=\frac{x^{t+1}\bar{\sigma}_i^{t+1}(h^t)}{\bar{\sigma}_i^{t+1}(h^t)+\sum_{j\neq i}\sigma_{j}^{t+1}(h^t)}+ v_{i,t+2}(\sigma|h_{dev}^{t+1}), \label{4.3x}
		\end{equation}
		where $h_{dev}^{t+1}$ is a successor of $h^t$ with the property that at battle $t+1$ player $i$ spent $\bar{\sigma}_i^{t+1}(h^t)$, and each player $j\neq i$ spent proportionally. And the expected payoff of player $i$ after history $h^t$ if she follows $\sigma$,
		\begin{equation} 
			v_{i,t+1}(\sigma|h^t)= \frac{x^{t+1}\sigma_i^{t+1}(h^t)}{\sum_{1\leq j \leq n}\sigma_{j}^{t+1}(h^t)}+ v_{i,t+2}(\sigma|h^{t+1}),
			\label{4.4x}
		\end{equation}
		where $h^{t+1}$ is a successor of $h^t$ with the property that at battle $t+1$, each player spent proportionally. For simplicity we take 
		\begin{equation}
			\frac{x^{t+1}+\ldots+x^{m}}{x^{t+1}}=k \notag,
		\end{equation}
		\begin{equation} B_i(h^t)=a,\notag\end{equation}
		\begin{equation}
			\sum_{j\neq i}B_{j}(h^t)=b,\notag\end{equation}
		\begin{equation}\bar{\sigma}_i^{t+1}(h^t)=\sigma_i^{t+1}(h^t)+\Delta=\frac{a}{k}+\Delta.\notag
		\end{equation} 
		where $\Delta$ is a real number. We can rewrite player $i$'s probability of winning battle $t+1$ if he plays $\bar{\sigma}_i^{t+1}(h^t)$ as
		\begin{equation}
			\frac{\bar{\sigma}_i^{t+1}(h^t)}{\bar{\sigma}_i^{t+1}(h^t)+\sum_{j\neq i}\sigma_{j}^{t+1}(h^t)}=\frac{\frac{a}{k}+\Delta}{\frac{a}{k}+\Delta+\frac{b}{k}}=\frac{a+\Delta k }{a+\Delta k + b}, \notag
		\end{equation} and player $i$'s probability of winning battle $t+1$ if he plays $\sigma_i^{t+1}(h^t)$ as
		\begin{equation}
			\frac{\sigma_i^{t+1}(h^t)}{\sum_{1\leq j\leq n}\sigma_{j}^{t+1}(h^t)}=\frac{\frac{a}{k}}{\frac{a}{k}+\frac{b}{k}}=\frac{a }{a+b}
			\notag.\end{equation}
		Since $\sigma$ is a proportional strategy profile, for any $t$, for any $h^t$, and for successor of histories where $h^{t+1}$ is a successor of $h^t$, $h^{t+2}$ is a successor of $h^{t+1}$, and so on up to and including $h^{m}$ is a successor of $h^{m-1}$, given that players follow proportional strategy profile, we have 
		\begin{equation}
			\frac{\sigma_i^{t+1}(h^t)}{\sum_{1\leq j\leq n}\sigma_{j}^{t+1}(h^t)}=\frac{\sigma_i^{t+2}(h^{t+1})}{\sum_{ 1\leq j\leq n}\sigma_{j}^{t+2}(h^{t+1})}=\ldots=\frac{\sigma_i^{m}(h^{m-1})}{\sum_{ 1\leq j\leq n}\sigma_{j}^{m}(h^{m-1})}=\frac{a}{a+b},\notag
		\end{equation} which means that player $i$ wins each battle after $h^t$ with equal probability if he/she follows $\sigma_i$. That is, if players follow the proportional strategy profile, the proportions of the remaining budgets stay constant throughout the battles. The same property satisfies for the strategy profile $(\bar{\sigma}_{i},\sigma_{-i})$ after history $h^{t+1}_{dev}$. Hence for successor of histories where $h^{t+2}_{dev}$ is a successor of $h^{t+1}_{dev}$, $h^{t+3}_{dev}$ is a successor of $h^{t+2}_{dev}$, and so on up to and including $h^{m}_{dev}$ is a successor of $h^{m-1}_{dev}$, given that players follow proportional strategy profile after history $h^{t+1}_{dev}$, we have 
		\begin{equation}
			\frac{\sigma_i^{t+2}(h^{t+1}_{dev})}{\sum_{ 1\leq j\leq n}\sigma_{j}^{t+2}(h^{t+1}_{dev})}=\frac{\sigma_i^{t+3}(h^{t+2}_{dev})}{\sum_{ 1\leq j\leq n}\sigma_{j}^{t+3}(h^{t+2}_{dev})}=\ldots= \frac{\sigma_i^{m}(h^{m-1}_{dev})}{\sum_{ 1\leq j\leq n}\sigma_{j}^{m}(h^{m-1}_{dev})}.\notag
		\end{equation}
		Now we can simply calculate player $i$'s probability of winning any battle after history $h^{t+1}_{dev}$, if player $i$ follows the strategy $\bar{\sigma_i}$
		\begin{equation}
			\frac{\sigma_i^{t+2}(h^{t+1}_{dev})}{\sum_{ 1\leq j\leq n}\sigma_{j}^{t+2}(h^{t+1}_{dev})}
			=\frac{a-\frac{a}{k}-\Delta}{a-\frac{a}{k}-\Delta+b-\frac{b}{k}}.\notag
		\end{equation}
		Therefore we can rewrite equation ($\ref{4.3x}$) as
		\begin{equation}
			v_{i,t+1}(\bar{\sigma}_i,\sigma_{-i}|h^t)=x^{t+1}\frac{a+\Delta k }{a+\Delta k + b}+ \frac{a-\frac{a}{k}-\Delta}{a-\frac{a}{k}-\Delta+b-\frac{b}{k}} (x^{t+2}+\ldots+x^{m}),\notag
		\end{equation}And we can rewrite equation ($\ref{4.4x}$) as
		\begin{equation}
			v_{i,t+1}(\sigma|h^t)= \frac{a}{a+b} (x^{t+1}+\ldots+x^{m}). \notag
		\end{equation}
		We show that $v_i(\sigma|h^t)-v_i(\bar{\sigma}_i,\sigma_{-i}|h^t)\geq0$, in other words show that
		\begin{equation} x^{t+1}(\frac{a}{a+b}-\frac{a+\Delta k }{a+\Delta k + b})+(x^{t+2}+\ldots+x^{m}) (\frac{a}{a+b}-\frac{a-\frac{a}{k}-\Delta}{a-\frac{a}{k}-\Delta+b-\frac{b}{k}})\geq0. \label{4.5xx}
		\end{equation}
		Since $k-1 = (x^{t+2}+\ldots+x^m)/(x^{t+1})$, we can rewrite inequality $(\ref{4.5xx})$ as
		\begin{equation} (\frac{a}{a+b}-\frac{a+\Delta k }{a+\Delta k + b})+(k-1) (\frac{a}{a+b}-\frac{a-\frac{a}{k}-\Delta}{a-\frac{a}{k}-\Delta+b-\frac{b}{k}})\geq0. \label{4.6xx}
		\end{equation}
		
		We can simplify inequality $(\ref{4.6xx})$ as 
		
		\begin{equation}
			\frac{b \Delta^2 k^3}{(a+b) (a (k-1)+b (k-1)-\Delta k) (a+b+\Delta k)} \geq 0 \label{newxx}.
		\end{equation}
		
		Inequality $(\ref{newxx})$ satisfies because we have the following conditions
		\begin{equation}
			a\geq \frac{a}{k}+\Delta, \notag
		\end{equation}
		\begin{equation}
			b(k-1)\geq0, \notag
		\end{equation}
		\begin{equation}
			\frac{a}{k}+\Delta\geq 0. \notag
		\end{equation}
		Thus, for any $\Delta$ we have $v_i(\sigma|h^t)-v_i(\bar{\sigma}_i,\sigma_{-i}|h^t)\geq0$.
	\end{proof}
	
	\section{Dynamic Blotto games with exogenous budget shocks and more general contest success functions}
	\label{sec:extension}
	
	In this section, we provide an extension of our main result to a more general setting with exogenous budget shocks and with any contest success function satisfying Skaperdas's (1996) axioms A1--A6.
	
	\noindent\textbf{Changes in the model}: A history of length $t \geq 1$ is a sequence
	\begin{equation}
		h^t :=(((w_1 ^{1},x_1 ^1),\ldots,(w_n ^{1},x_n ^1)),\dots,((w_1 ^{t},x_1 ^t),\ldots,(w_n ^{t},x_n ^t)))
	\end{equation}
	satisfying the following conditions
	\begin{itemize}
		\item[(i)] For each $1 \leq i \leq n$ and for each $1 \leq t'\leq t$,  $w_i^{t'} \in [0,W_i+ z^{t'}_i-\sum_{j< t'}w_i ^{j}]$, where $z^{t'}_i\in \mathbb{R}$ represents the \textit{exogenous} budget shock player $i$ receives, if any, before battle $t'$ takes place. The amount of the exogenous shock $z^{t'}_i$ is announced publicly to all players before battle $t'$.
		\item[(ii)] For each battle $t'\leq t$, there exists a unique player $i$ such that $x_i^{t'} = x^{t'}$ and for all $j \neq i$, $x_j^{t'}=0$.
	\end{itemize}
	Part (i) introduces a new variable $z^{t'}_i$ which needs more eloboration. Note that everyone receives the news about the exogenous budget shock at the same time and before the next battle takes place, so the dynamic game is still a game under complete information. In the context of sequential elections, an example would be that a private donor decides to contribute resources to the campaign of a (presidential) candidate after several mini-elections have taken place.
	
	Accordingly, the remaining budget of player $i$ after history $h^t\in H$ is defined as $\hat{B}_i(h^t)= \max \{W_i +\sum_{j\leq t+1} z_i ^{j}-\sum_{j\leq t} w_i ^{j},0\}$ where for every $j\leq t$, $w_i ^{j}$ is a realized spending of history $h^{t}$.  For player $i$, a pure strategy $\sigma_i$ is a sequence of $\sigma_i^{t}$'s such that for each $t$, $\sigma_i^t$ assigns, to every $h^{t-1} \in H^{t-1}$, allocation $\sigma_i^t(h^{t-1}) \in [0,\hat{B}_i(h^{t-1})]$. 
	
	\noindent\textbf{Proportional strategy profile:} For any $t$, for any nonterminal history $h^{t-1}\in H-\overline{H}$, and for any player $i$, let
	\begin{equation}
		\label{proportionality}
		\sigma_i^{t}(h^{t-1})= \hat{B}_i(h^{t-1}) \frac{x^{t}}{x^{t}+\ldots+x^{m}}.
	\end{equation} Note that for any subgame starting at history $h^{t-1}$ and for any player $i$, the knowledge of player $i$ on the variable budget set is $\{z_1^t,\ldots,z_n^{t}\}$. Therefore the strategy $\sigma$ is defined on $\hat{B}_i(h^t)$ but not on future variable budgets, which is exogenously given each time before a battle takes place. We call $\sigma$ the proportional (pure) strategy profile. Note that under $\sigma$, no matter what the other players do, every player proportionally allocates her available budget over the remaining battles. 
	
	Mutatis mutandis, the rest of the model remains the same as the model presented in section~\ref{sec:model}.
	
	We now state our second main result, which extends Theorem~\ref{thm2} to the case with variable budgets and more general contest success functions.
	
	\begin{myth}[General existence and characterization]
		\label{thm3}
		For any $n$-player dynamic contest with variable budgets, the following proportional strategy profile, $\sigma$, is a subgame perfect equilibrium. For any $t$, for any nonterminal history $h^{t-1}\in H-\overline{H}$, and for any player $i$, 
		\begin{equation}
		\label{proportionality}
		\sigma_i^{t}(h^{t-1})= \hat{B}_i(h^{t-1}) \frac{x^{t}}{x^{t}+\ldots+x^{m}}.
		\end{equation}
	\end{myth}
	\begin{proof}
		First, we show that a proportional strategy profile $\sigma^* \in \Sigma$ in the dynamic contest is a Nash equilibrium by proving that for any player $i$ a strategy $\sigma_i \in \Sigma_i$ is a best response to $\sigma^{*}_{-i}$ whenever $\sigma_i = \sigma^*_i$ with fixed budgets, and then we extend this result to contests with variable budgets.
		Let $w$ be the spending sequence associated with $(\sigma_i, \sigma^*_{-i})$ and  $w_{-i}=(w^t_{-i})_{t\leq m}$ denote the spending sequence excluding player $i$, where for all $t$, $w^t_{-i}=(w^t_1,\ldots,w^t_{i-1}, w^t_{i+1}, \ldots, w^t_n)$.%
		\footnote{To be sure, one may condition his strategy on the winners of the previous battles and also on the previous battle spendings. However, without loss of generality, we can confine attention to the spending sequence, $w$, that is associated with the given strategy profile, because the payoff received from the previous battles does not affect the payoff that can be received from the remaining ones as the payoff function is additive.}
		We show that 
		\begin{equation}
			\sigma_i\in\argmax_{\sigma_{i}'}v_{i}(\sigma^*_{-i},\sigma_{i}'| \o),
		\end{equation}
		that is, player $i$'s best response to $\sigma^*_{-i}$ associated with $w_{-i}$ is $\sigma_i$ associated with $w_i$. Given that all players but $i$ follow the spending sequence $w_{-i}$, player $i$'s expected prize from a $1$-prize battle $t_1$ for any $w_{i}^{t_1}$, which we treat as a variable, is given by
		\begin{equation}
			\label{eq:one-vote}
			\frac{\beta(w_{i}^{t_1})^{\alpha}}{(\beta(w_{i}^{t_1})^{\alpha}+\beta\sum_{j\neq i}(w_{j}^{t_1})^{\alpha})}.
		\end{equation}
		Differentiating (\ref{eq:one-vote}) with respect to $w_{i}^{t_1}$ gives
		\begin{equation}
			\label{eq:one-vote-margin}
			\displaystyle\frac{\alpha(w_{i}^{t_1})^{\alpha-1}\sum_{j\neq i}(w_{j}^{t_1})^{\alpha}}{((w_{i}^{t_1})^{\alpha}+\sum_{j\neq i}(w_{j}^{t_1})^{\alpha})^{2}},
		\end{equation}%
		which is player $i$'s marginal gain from the battle $t_1$. We next consider
		a $k$-prize battle $t_k$ for some $k\in\{1,...,m\}$. By our supposition each player except player $i$
		spends in proportion to the prize of the battle, i.e., for each $j \neq i
		$, $w_{j}^{t_k}=kw_{j}^{t_1}$. We next show that player $i$'s best response to proportional allocation is also to spend
		in proportion to the prize at battle $t_k$, i.e., $w_{i}^{t_k}=kw_{i}^{t_1}$. In this case,
		player $i$'s expected prize for any $w_{i}^{t_k}$ from $k$-prize battle $t_k$ is
		\begin{equation}
			\label{eq:k-vote}
			\dfrac{k \beta (w_{i}^{t_k})^\alpha}{(\beta (w_{i}^{t_k})^\alpha+\beta\sum_{j\neq i}w_{j}^{t_k})}=\dfrac{%
				k(w_{i}^{t_k})^\alpha}{((w_{i}^{t_k})^\alpha+k^\alpha\sum_{j\neq i}(w_{j}^{t_1})^\alpha)}.
		\end{equation}%
		Differentiating (\ref{eq:k-vote}) with respect to $w_{i}^{t_k}$
		gives
		\begin{equation}
			\label{eq:k-vote_margin}
			\dfrac{\alpha k^{\alpha+1} (w_{i}^{t_k})^{\alpha-1} \sum_{j\neq i}(w_{j}^{t_1})^\alpha}{((w_{i}^{t_k})^\alpha + k^\alpha\sum_{j\neq i}(w_{j}^{t_1})^\alpha)^2},
		\end{equation}%
		which is player $i$'s marginal gain from the $k$-prize battle $t_k$.
		Next, we show that for $w_{i}^{t_k}=kw_{i}^{t_1}$, Expression~\ref{eq:k-vote_margin} equals Expression~\ref{eq:one-vote-margin}. First, Expression~\ref{eq:k-vote_margin} equals
		\begin{equation}
			\dfrac{\alpha k^{\alpha+1} (w_{i}^{t_k})^{\alpha-1} \sum_{j\neq i}(w_{j}^{t_1})^\alpha}{((w_{i}^{t_k})^\alpha + k^\alpha\sum_{j\neq i}(w_{j}^{t_1})^\alpha)^2}	
			=
			\dfrac{\alpha k^{\alpha+1} (kw_{i}^{t_1})^{\alpha-1} \sum_{j\neq i}(w_{j}^{t_1})^\alpha}{((kw_{i}^{t_1})^\alpha + k^\alpha\sum_{j\neq i}(w_{j}^{t_1})^\alpha)^2}.
		\end{equation}
		Cancelling out $k$'s leads to
		\begin{equation}
			\displaystyle\frac{\alpha(w_{i}^{t_1})^{\alpha-1}\sum_{j\neq i}(w_{j}^{t_1})^{\alpha}}{((w_{i}^{t_1})^{\alpha}+\sum_{j\neq i}(w_{j}^{t_1})^{\alpha})^{2}},
		\end{equation} 
		which is Expression~\ref{eq:one-vote-margin}.
		We showed that if player $i$ allocates proportionally to $k$-prize battle, then his marginal gain from that battle is equal to his marginal gain from $1$-prize battle provided that others allocate proportionally. Thus, there is no incentive for player $i$ to deviate from proportional allocation when others allocate proportionally. Hence, the proportional strategy profile $\sigma$ is a Nash equilibrium of the dynamic contest.\par
		
		Now we show that $\sigma^*$ is a subgame perfect equilibrium. In other words, for every $h \in H,$ $ \sigma^*$ induces an equilibrium in the subgame starting with history $h$. By definition, the subgame starting with history $h$ is a game (i.e., dynamic contest) and $(\sigma^*| h)$ is a proportional strategy profile. Thus, by an analogous argument used in the first part of the proof, $(\sigma^*| h)$ is a Nash equilibrium in the subgame starting with history $h$.%
		\footnote{Note that in the first part we showed that a  proportional strategy profile is a Nash equilibrium in any dynamic contest.}
		That is, we obtain for every $i$ and every $h$
		\begin{equation}
			(\sigma^*_{i}| h)\in\argmax_{\sigma_{i}'}v_{i}(\sigma^*_{-i},\sigma_{i}'| h),
		\end{equation}	
		Therefore, $\sigma^*$ is a subgame perfect equilibrium, so the dynamic contest satisfies  proportionality.
		
		So far, we have shown that in the beginning of the dynamic contest with fixed budgets players allocating their budgets proportionally is a subgame perfect Nash equilibirum of the dynamic contest. 
		It implies that in the dynamic contest with variable budgets, players allocating their budgets proportionally is also a Nash equilibirum in the beginning of the game, because the initial budgets and payoffs are the same in the begining. After the competition in the first battle, players' budgets are updated, which defines a new dynamic contest. Applying the same argument again on this game shows that proportional allocation is a Nash equilibrium in this subgame as well. By repeating this process, we conclude that there is a subgame perfet equilibrium in which players allocate their budgets proportionally to each battle however the budgets might evolve throughout the contest.
	\end{proof}
	
	\section{Discussion and concluding remarks}
	
	In this paper, we have introduced a general model of dynamic $n$-player Blotto games. In this framework, the budgets and battlefields are possibly asymmetric, and there are no restrictions on the number of players or the number of battlefields (e.g., odd or even). In this context, we studied the proportional allocation of resources and the equilibrium behavior. We showed that the strategy profile in which players allocate their resources proportionally at every history is a subgame perfect equilibrium.  We have shown that our results are robust to exogenous shocks on budgets. Moreover, the results do not depend on the specific contest success function used in the competition as long as it satisfies Skaperdas's (1996) axioms (A1--A6). An open question for future research is whether the proportional strategy profile is the unique subgame perfect equilibrium or not in $n$-player dynamic contests. 
	
	
	As we have mentioned, Blotto games can be applied to many economic and political situations as Borel (1921) himself envisioned. As an example, consider sequential elections as an $n$-player dynamic multi-battle contest where political candidates choose how they distribute their limited resources over multiple ``battlefields'' or states as in the U.S. presidential primaries. In this context, our results imply that proportionality is immediately rectified once one has candidates who maximize their electoral vote instead of simply maximizing their probability of winning, despite the presence of the winner-take-all feature.
	
	To achieve proportionality in at least the U.S. presidential primaries which
	have the winner-take-all feature, a viable policy suggestion could be to
	provide additional incentives for players to induce them to win as many
	delegates as possible in the entire presidential primaries overall. For
	instance, the electoral system may provide them additional funding in the
	ensuing presidential race where these incentives are positively linked to
	the number of delegates won by the presidential player in the primaries.
	Such incentives can be very effective at the margin. As a matter of fact,
	even in the absence of any such additional pecuniary incentives, for one
	reason or another, players seem to already have the behavioral trait of
	maximizing their expected delegates themselves and do not want to stop
	pumping campaign funding to their remaining primaries even though they have
	already guaranteed winning the majority of the delegates in the primaries.
	The main reason that the players may try win more delegates beyond what they
	would need to guarantee their presidential candidacy (i.e., the main reason
	that they still might keep investing in the remaining primaries even though
	they know that it will not affect their chances of winning any further
	delegates) could be that they care about entering the U.S. presidential race
	with an impressive momentum gained in the presidential primaries, as Hillary
	Clinton tried to do against the late surge of Bernie Sanders in the U.S.
	Democratic primaries in 2016 even though she had already accumulated more
	than sufficiently many delegates to win her party's presidential candidacy
	up to that point. Nevertheless, to ensure proportionality, the parties or
	the electoral system may consider boosting players' tendency to maximize
	their expected delegates via some additional pecuniary incentives, which may
	help at the margin at least for the players who may simply try to maximize
	their winning probability in their U.S. presidential primaries.
	
	\bibliographystyle{chicago}

	\nocite{friedman1958}
	\nocite{osorio2013}
	
	\nocite{duggan2007}
	\nocite{barelli2014}
	\nocite{laslier2002}
	
	\nocite{thomas2017}
	\nocite{montero2016}

	\nocite{acharya2021}
	\nocite{arad2012}
	\nocite{konrad2009}
	\nocite{konrad2018}
	\nocite{konrad2009book}
	\nocite{klumpp2019}
	\nocite{nagel1995}
	\nocite{hinnosaar2018}
	\nocite{roberson2006}
	\nocite{kvasov2007}
	\nocite{deck2012}
	\nocite{rinott2012}
	\nocite{brams1973}
	\nocite{brams1974}
	\nocite{brams1982}
	\nocite{borel1921}
	\nocite{deck2017}
	\nocite{dziubinski2021}
	\nocite{duffy2015}
	\nocite{duffy2017}
	\nocite{ewerhart2019}
	\nocite{fu2015}
	\nocite{harris1985}
	\nocite{klumpp2006}
	\nocite{kovenock2009}
	\nocite{krueger1974}
	\nocite{lake1979}
	\nocite{sela2013}
	\nocite{skaperdas1996}
	\nocite{stahl1993}
	\nocite{strumpf2002}
	\nocite{tullock1967}
	\nocite{tullock1974}
	\nocite{li2021}
	
	\bibliography{primaries}

\end{document}